# On-off switchable nonreciprocal negative refraction in non-Hermitian photon-magnon hybrid systems

Junyoung Kim[†], Bosung Kim[†‡], Bo-Jong Kim, Haechan Jeon, and Sang-Koog Kim[*]

National Creative Research Initiative Center for Spin Dynamics and Spin-Wave Devices, Nanospinics Laboratory, Research Institute of Advanced Materials, Department of Materials Science and Engineering, Seoul National University; Seoul 151-744, Republic of Korea

[†]These authors contributed equally to this work

[‡]Present address: SK Hynix Inc., Icheon 17336, Republic of Korea

[*]Corresponding author. Email: sangkoog@snu.ac.kr

**Abstract**

Photon-magnon coupling, where electromagnetic waves interact with spin waves, and negative refraction, which bends the direction of electromagnetic waves unnaturally, constitute critical foundations and advancements in the realms of optics, spintronics, and quantum information technology. Here, we explore a magnetic-field-controlled, on-off switchable, nonreciprocal negative refraction within a non-Hermitian photon-magnon hybrid system. By integrating an yttrium iron garnet film with an inverted split-ring resonator, we discover pronounced negative refraction driven by the system's non-Hermitian properties. This phenomenon exhibits unique nonreciprocal behavior dependent on the signal's propagation direction. Our analytical model sheds light on the crucial interplay between coherent and dissipative coupling, significantly altering permittivity and permeability's imaginary components, crucial for negative refraction's emergence. This work pioneers new avenues for employing negative refraction in photon-magnon hybrid systems, signaling substantial advancements in quantum hybrid systems.

**Introduction**

In the advancing fields of optics, magnonics, and quantum information science, the interaction between photons and magnons within photon-magnon hybrid (PMH) systems and associated non-Hermitian dynamics are at the forefront of innovative research [1-8]. These systems reveal new complexities in optical and magnetic and their coupled behaviors and have been of a broad interest owing to their potential applications in quantum information devices such as quantum networks [9], transducers [10-12], memories [13], entanglement [14,15], and sensors [16,17]. Central to these investigations is the concept of non-Hermiticity, marked by an imaginary component in the coupling constant [6,18-20], leading to novel behaviors in non-reciprocal signal transmission, reflection, and absorption [7,8,21,22] that challenge the conventional Hermitian principles.

Our groundbreaking study introduces unprecedented insights into the negative refractive index (NRI) phenomenon within a PMH system. NRI, a counterintuitive phenomenon where electromagnetic waves bend inversely in a medium [23-26]. Previously recognized in optics and magnonics, the occurrence and mechanisms of NRI in PMH systems have been unexplored. Our work represents the first experimental demonstration of magnetic-field controllable, non-reciprocal NRI in PMH systems, uncovering its inherent non-reciprocal nature associated with non-Hermiticity.

To elucidate the observed robust negative refraction phenomenon in PMH systems, we employed an analytical circuit model that connects NRI with the non-Hermitian aspects of PMC systems. This model, considering the balance between energy loss and gain reflected in both real and imaginary components of the coupling constant, is pivotal to understanding NRI's occurrence. This work not only offers new insights into PMC and NRI but also underscores the importance of non-Hermitian dynamics in the advancement of photonics and magnonics, indicating a significant leap in optical and magnonic quantum hybrid technologies.

**Experimental results: Observation of NRI**

Figure 1a shows our measurement setup for the scattering parameters $S_{ij}$ as functions of both the AC current frequency ($f_{AC}$) and the static magnetic field ($\mu_0 H$), as well as the geometry of our PMH sample. The sample is composed of an yttrium iron garnet (YIG) film, and an inverted split-ring resonator (ISRR) patterned on ground plane. Optical images of the sample's front and back sides are displaced in Fig. 1b. The magnitude (first row) and phase (middle) of the experimentally measured $S_{21}$ (from port 1 to port 2) and $S_{12}$ (from port 2 to port 1) spectra are given in Fig. 1c. Owing to the sufficient-strength coupling between the YIG's ferromagnetic resonance (FMR) mode and the ISRR's photon mode, two split (higher- and lower-branch) hybrid modes appear in the |$S_{21}$| and /$S_{12}$| spectra, but with opposite asymmetry between /$S_{21}$/ and /$S_{12}$| (see the varying contrasts denoted by the blue color around the coupling center in the field). These split hybrid modes represent anti-crossing dispersion at or near the common resonance frequency, as previously reported [27]. Note that all the spectra shown in Fig. 1c were obtained by subtracting the background signal of alone the microstrip line from that of the entire structure to investigate the interest region only [28] (for details see supplementary Note 3). In the magnitude and phase spectra, the intersecting solid black lines denote the uncoupled individual photon ($\widetilde{\omega}_{ISRR}$) and magnon ($\widetilde{\omega}_r$) modes versus $\mu_0 H_{DC}$, while the dashed lines correspond to the real component of the eigenvalues for their coupled modes, as expressed by:

$$\widetilde{\omega}_{\pm} = \frac{1}{2}\{(\widetilde{\omega}_{ISRR} + \widetilde{\omega}_r) \pm \sqrt{(\widetilde{\omega}_{ISRR} - \widetilde{\omega}_r)^2 + (\omega_{ISRR}\kappa)^2}\}. \tag{1}$$

Here, $\kappa$ represents the coupling constant, characterized as a complex value [19,20,27]. The resonance angular frequencies are defined as $\widetilde{\omega}_{ISRR} = \omega_{ISRR} - i \cdot \Delta\omega_{ISRR}$ and $\widetilde{\omega}_r = \omega_r - i \cdot \Delta\omega_r$, where $\omega_{ISRR}$ and $\omega_r$ correspond to the resonance frequencies, and $\Delta\omega_{ISRR}$ and $\Delta\omega_r$ to the linewidths of each mode (Supplementary Notes 1 and 2). The observed anti-crossing (or level repulsion) dispersion profiles predominantly arise from coherent coupling, as evidenced by the real component of $\kappa$, which is fitted as Re[$\kappa$] ~ 0.0296 using Eq. 1. On the other hand, the distinct asymmetries present in the profiles of the higher- and lower-branch modes, particularly when comparing $|S_{21}|$ and $|S_{12}|$ are attributed to the interaction between coherent and dissipative coupling, breaking time-reversal symmetry and thus leading to nonreciprocal signal transmission behaviors, as addressed in previous works [21,22,29]. This dissipative coupling, mediated by traveling waves in the microstrip line [30], is characterized by the imaginary component of $\kappa$, with fitted values indicating Im[$\kappa$] at $+0.0088i$ for $|S_{21}|$ and $-0.0088i$ for $|S_{12}|$, where the sign change is ascribed to a $\pi$ phase difference in the traveling waves between the $S_{21}$ and $S_{12}$. Further discussion will be addressed in the 'NRI mediated by non-Hermiticity' section.

The refractive index $n$ of our hybrid sample, which exhibits significant PMC characteristics, was determined by analyzing the measured $S_{ij}$ data, in accordance with Eq. 2 [28,31-33]:

$$n = n' - i \cdot n'' = -\frac{1}{ik_0 l_s} \ln P, \qquad (2)$$

where $P$ is the time-independent wave function for microwaves propagating in the ISRR/YIG hybrid sample, $l_s$ is the effective sample length, and $k_0$ the electromagnetic wave number in vacuum. $P$ is given as $\frac{\bar{S}+S_{21}-\Gamma}{1-(\bar{S}+S_{21})\Gamma}$ with $\Gamma = \frac{z-1}{z+1}$ the reflection coefficient at the interface between the ISRR/YIG hybrid structure and the microstrip line. $z$, the normalized impedance of the

ISRR/YIG hybrid structure to 50 Ω, is given as $z = \sqrt{\frac{(1+\bar{S})^2-(S_{21})^2}{(1-\bar{S})^2-(S_{21})^2}}$). The relationships of $\varepsilon_{eff} = n/z$ and $\mu_{eff} = n \cdot z$ allow for the determination of the effective relative permittivity $\varepsilon_{eff}$ and permeability $\mu_{eff}$, with comprehensive results available in Fig. S5 (for further details, see Supplementary Notes 3).

In Fig. 1c, the third row illustrates the real components of the refractive index ($n'$) for the YIG/ISRR hybrid sample. A distinct area, where $n'$ manifests predominantly negative values, is identifiable in the anti-crossing region, indicated by a blue color along the higher-branch mode for $S_{21}$ and the lower-branch mode for $S_{12}$. These $n'$ values reach as low as -9.0 and -9.7 for $S_{21}$ and $S_{12}$, respectively, from an initial uncoupled state value of $n' = +0.2$. The NRI spans a specific magnetic field region from $\mu_0 H = 50.3$ to 64.4 mT for $S_{21}$ and from $\mu_0 H = 56.5$ to 70.7 mT for $S_{12}$, extending over a wide frequency range of $\Delta f_{AC} = 412$ MHz and a field range of $\Delta \mu_0 H = 19.9$ mT for $S_{21}$, and $\Delta f_{AC} = 498$ MHz and $\Delta \mu_0 H = 22.3$ mT for $S_{12}$. Notably, the NRI characteristics resulting from the $S_{12}$ and $S_{21}$ measurements differ in their operating areas, demonstrating non-overlapping and functionally valuable nonreciprocal properties for opposite input-output signal directions.

To validate the optical prerequisites essential for NRI and elucidate the PMC-linked NRI mechanism, we revisited the NRI optical criteria. Contrary to the standard requirement for lossless materials, which mandates both $\varepsilon'_{eff} < 0$ and $\mu'_{eff} < 0$ (double-negative condition) [23-25], our observations suggest an alternative mechanism for lossy materials. This mechanism adheres to the generalized optical condition: $\varepsilon'_{eff} \cdot \mu''_{eff} + \varepsilon''_{eff} \cdot \mu'_{eff} < 0$, a prerequisite for backward phase velocity and hence indicative of NRI [34-39]. The calculations, shown in the last row of Fig. 1c, based on both the real and imaginary components of $\varepsilon_{eff}$ and $\mu_{eff}$ (Fig. S5), demonstrate negative values (indicated by blue color) coinciding with regions where $n' < 0$, thereby validating this generalized condition. The positive $\varepsilon''_{eff}$ and negative $\mu''_{eff}$ values observed in the NRI region, as shown in Fig. S5, confirm that PMC leads to sign changes in the electric loss ($\varepsilon''_{eff}$) and magnetic loss ($\mu''_{eff}$) terms, through energy exchange between the YIG and ISRR resonators. This process aligns with the generalized optical condition necessary for achieving NRI.

**NRI mediated by non-Hermiticity.**

The emergence of NRI in specific field regions is closely associated with the interplay of energy gain and loss in a non-Hermitian PMH system. Our analysis of the linewidths of the two distinct hybrid modes, derived from the $|S_{21}|$ and $|S_{12}|$ spectra (Fig. 2a), reveals that at particular field strengths, denoted as the lower field $H_-$ and higher field $H_+$, exceptionally narrow linewidths are observed exclusively in either the higher- or lower-branch modes: $H_-$ = 50.3 and $H_+$ = 64.4 mT for $|S_{21}|$, and $H_-$ = 56.5 and $H_+$ = 70.7 mT for $|S_{12}|$. Such narrow linewidths are indicative of zero damping, a distinct characteristic of non-Hermitian systems [6,21,22,40,41]. Zero damping represents a transition state, where damping alternates between negative and positive values.

To further verify zero and negative damping states, we plotted the real and imaginary components of the hybrid modes' eigenvalues (Fig. 2b), employing Eq. (1) for both $S_{21}$ and $S_{12}$, with fitted values of $\kappa = 0.0296 + 0.0088i$ and $\kappa^* = 0.0296 - 0.0088i$, respectively. This approach, which involved extracting intrinsic values from each mode and fitting them to the experimental data (Fig. 2b), revealed that there exist two zero-damping points at the marked $H_-$ and $H_+$ values in the linewidth ($\Delta\omega$) profiles for the higher-branch mode in $S_{21}$ and the lower-branch mode in $S_{12}$. Between these two fields, all $\Delta\omega$ values are negative (indicated by the blue shaded area), denoting anti-damping zones, which coincide with the field regions, where experimentally observed NRI occurs, as shown in the top panel of Fig 2b. These anti-damping regions, emerging from non-Hermitian dissipative coupling [18-22], coincide with transitions in the damping term from positive to negative. Correspondingly, the loss terms $\varepsilon''_{eff}$ and $\mu''_{eff}$ alter their signs within this anti-damping region, with $\varepsilon''_{eff}$ becoming negative and $\mu''_{eff}$ turning positive, as detailed in Fig. S5. Consequently, the generalized optical criterion for NRI, $\varepsilon'_{eff} \cdot \mu''_{eff} + \varepsilon''_{eff} \cdot \mu'_{eff} < 0$, is intrinsically associated to the anti-damping phenomenon in the non-Hermitian nature in our YIG/ISRR hybrid sample.

**Effect of PMC strength on NRI and operating area**

To explore further how the occurrence of NRI varies with the PMC strength, we experimentally varied the coupling strength. This manipulation was achieved by inserting scotch tapes with different thicknesses ($d$), ranging from $d = 0$ to 3.0 mm, between the YIG and the microstrip line. The experimental data for $|S_{21}|$, $|S_{12}|$, and the calculated values of the corresponding $n'$ for $d = 0$, 0.2, 0.8, and 3.0 mm are depicted in Fig. 3. As $d$ increases, resulting in a reduction in coupling strength $\kappa/\kappa_c$ (where $\kappa_c$ represents the coupling constant at $d = 0$), the region exhibiting negative $n'$ values diminishes, visually represented by shrinking blue-colored areas. Notably, at $\kappa/\kappa_c = 0.5$ (corresponding to $d = 0.8$ mm), NRI eventually disappears. These specific coupling strength values were determined by fitting to the higher- and lower- branch modes, as indicated by the dotted lines in Fig. 3. At $d = 3$ mm, the individual photon and magnon modes (dotted lines) are clearly separated, intersecting at a common resonant frequency. However, the zoomed insets in the left column of Fig. 3 still exhibit anti-crossing in the dispersion spectra, indicating persistent coupling with a non-zero value of $\kappa/\kappa_c = 0.1$. These experimental observations underscore the importance of significant PMC strength for enabling NRI in the hybrid system and establish a threshold coupling strength necessary for NRI manifestation.

**Analytical Circuit Model**

To enhance our understanding of non-Hermitian PMC associated NRI, quantitatively reproduce the experimentally observed NRI in the PMC region, and identify key controllable parameters, we developed an analytical circuit model based on transmission line theory (Fig. 4a) [42-44]. This model incorporates the interactive dynamics between the YIG and ISRR resonators, as well as the microstrip line responsible for pumping and detecting AC transmission signals. The YIG's excitation and its coupling with the AC current in the microstrip line (governed by Ampere's law), are quantified by a coupling constant parameter $M_0$. The microstrip line's effective inductance and capacitance, isolated from the ISRR through an insulating substrate, are denoted by $L_0$ and $C_0$, respectively. The ISRR's effective inductance, capacitance, and resistance are expressed as $L_{ISRR}$, $C_{ISRR}$, and $R_{ISRR}$. Additionally, the YIG's effective inductance is expressed as $L_{YIG} = L_0 \left( \frac{\omega_m(\omega_H+\omega_m+i\omega\alpha)}{\omega_r^2-\omega^2+i\omega\alpha(2\omega_H+\omega_m)} \right)$, as derived from the Landau-Lifshitz-Gilbert (LLG) equation. Here, $\omega_H = \gamma_0\mu_0 H$ is the angular frequency of AC magnetic field, $\omega_m = \gamma_0\mu_0 M_s$ is the effective characteristic frequency of magnetization, with $\omega_r^2 = \omega_H(\omega_H + \omega_m)$ the FMR frequency, and $\alpha$ the FMR damping constant [45-47]. The inductance of YIG, when coupled with microstrip line, is modified to $L_{YIG,ext}$

In the circuit model given in Fig. 4a, the hybrid system comprising YIG, ISRR, and microstrip is represented by integrating the per-unit-length series impedance ($z_{se}$) and shunt admittance ($y_{sh}$), incorporating the circuit parameters previously outlined [48,49]:

$$z_{se} \cdot l_s = i\omega L_0 + i\omega M_0^2 L_{YIG,ext}, \qquad (3)$$

$$y_{sh} \cdot l_s = \left[ \frac{1}{i\omega C_0} + \left( \frac{1}{i\omega L_{ISRR}+i\omega M_c^2 L_{YIG}} + i\omega C_{ISRR} + \frac{1}{R_{ISRR}} \right)^{-1} \right]^{-1}, \qquad (4)$$

where $\omega$ is the angular frequency of AC currents flowing along the microstrip line, and $l_s$ (= 0.005 $m$) is the length of the sample. The parameters of $n$, $\varepsilon_{eff}$, $\mu_{eff}$, and $z$ are reformulated as [48-50],

$$n = \frac{\sqrt{z_{se}y_{sh}}}{i \cdot k_0} \left[\frac{\frac{\theta}{2}}{\sin\left(\frac{\theta}{2}\right)}\right] = \frac{\theta}{l_s \cdot k_0}, \quad (5)$$

$$\varepsilon_{eff} = \frac{y_{sh} \cdot G}{i \cdot \omega \cdot \varepsilon_0} \left[\frac{\frac{\theta}{2}}{\sin\left(\frac{\theta}{2}\right) \cdot \cos\left(\frac{\theta}{2}\right)}\right], \quad (6)$$

$$\mu_{eff} = \frac{z_{se}}{i \cdot \omega \cdot \mu_0 \cdot G} \left[\frac{\left(\frac{\theta}{2}\right) \cdot \cos\left(\frac{\theta}{2}\right)}{\sin\left(\frac{\theta}{2}\right)}\right]. \quad (7)$$

By inputting numerical values obtained from the scattering parameters measured directly for each of the ISRR and YIG, along with the fitted values of the mutual coupling constants, $M_0 = 0.085$ and $M_c = 0.0093 - 0.0028i$ for $S_{21}$ ($M^*_c = 0.0093 + 0.0028i$ for $S_{12}$), into Eqs. 3-7, we were able to numerically calculate the analytical model for $|S_{21}|$ and $|S_{12}|$, as well as the corresponding $n'$ (Fig. 4b). Note that we employed the fitted mutual coupling constant $M^*_c = 0.0093 + 0.0028i$ for $S_{12}$, reflecting a sign inversion in the imaginary component of $M_c$ for $S_{21}$. The difference between the coupling constants $\kappa_c$ observed in the experimental data and $M_c$ used in the analytical circuit model can be attributed to the circuit model's incorporation of mutual inductance. This inclusion results in variations in $M_c$ depending on the inductance level of the utilized circuit components. However, the absolute ratio of the real to imaginary components of $\kappa_c$ matches the corresponding ratio of $M_c$ applied in the circuit model. Figure 4b reveals that the negative values of $n'$ and the operating area and its position, highlighted in blue color along the higher- and lower-branch modes for $S_{21}$ and $S_{12}$ accurately reproduce the experimentally observed NRI regions as shown in the third row of Fig. 1c and at the top of Fig. 2b. Additionally, the analytical circuit model enabled us to calculate the real and imaginary parts of $\varepsilon_{eff}$ and $\mu_{eff}$, along with the generalized optical condition for NRI. The results of these calculations mirror trends observed in the experimental data (see Fig. S8), indicating that the model effectively captures the essential physics underlying the hybrid system and its NRI phenomena.

To establish a correlation between the non-reciprocal characteristics of the experimentally observed NRI, including its operating region and position, and the magnitude and sign of the imaginary component of $M_c$, we conducted further calculations of $|S_{21}|$, $|S_{12}|$, and the corresponding $n'$ values with reduced magnitudes of $|\text{Im}[M_c]|$, such as 0.0024, 0.002, and 0, while maintaining its positive real counterpart, $\text{Re}[M_c] = 0.0093$, for both $M_c$ and $M^*_c$ (Fig. 4c). As the real part of $M_c$ remains unchanged across all calculations, the dispersion curves of the higher- and lower-branch modes including their opening gap largely remain unaffected by variations in the magnitude and sign of $\text{Im}[M_c]$. Nonetheless, the asymmetry in magnitude between the lower and higher modes intensifies with an increase in $|\text{Im}[M_c]|$. Most strikingly, the area exhibiting negative refraction narrows as $|\text{Im}[M_c]|$ decreases. With $\text{Im}[M_c] = 0$ (indicating only coherent coupling), NRI vanishes, even in the presence of substantial coupling strength. Furthermore, the sign of $\text{Im}[M_c]$ specifically alters the NRI operational position, manifesting non-reciprocal behavior based on the direction of the input-output signal. Therefore, it become clear that both the sign and magnitude of the imaginary coupling constant component, indicative of the non-Hermitian nature, plays a crucial role in the emergence of NRI, determining the extent of its operating area and its non-reciprocal behavior.

In Fig. 4d, we plotted the field range where NRI is found, along with the peak value of $n'$ as a function of $\kappa/\kappa_c$ ($\kappa^*/\kappa^*_c$) within the range of 0 to 1.5 at an increment of 0.1. These calculations (solid and dashed lines) were conducted with $M_c = 0.0093 - 0.0028i$ for $S_{21}$ ($M^*_c = 0.0093 + 0.0028i$ for $S_{12}$). The analytical calculations are in quantitative agreement with the corresponding experimental data (denoted by symbols) for various $\kappa$ values with constant values of $\kappa_c = 0.0296 + 0.0088i$ for S$_{21}$ and $\kappa_c^* = 0.0296 - 0.0088i$ for S$_{12}$. Below $\kappa/\kappa_c$ ($\kappa^*/\kappa_c^*$) = 0.5, the field range for NRI does not exist. However, as $\kappa/\kappa_c$ ($\kappa^*/\kappa_c^*$) exceeds 0.6, $n'$ suddenly transitions to large negative values ranging from -8.7 to -10.4 for $0.7 < \kappa/\kappa_c$ ($\kappa^*/\kappa_c^*$) $<$ 1.5. As mentioned earlier, the field range is closely related to the anti-damping region, $\Delta H_{ND}$, identified between two different fields exhibiting zero damping. For $\kappa/\kappa_c$ ($\kappa^*/\kappa_c^*$) > 0.6, two zero damping field points emerge, and the anti-damping range expands with increasing $\kappa/\kappa_c$. These analytical findings confirm that the NRI behaviors observed in the YIG/ISRR hybrid sample are significantly influenced by the non-Hermitian nature of our system. This influence is associated with dissipative coupling and its strength, as represented by the imaginary value of the coupling constant. The interplay between coherent and dissipative coupling is essential to the manifestation of characteristic nonreciprocal NRI behaviors.

**Discussion**

Our study represents a significant leap forward in PMC technologies, demonstrating remarkable non-reciprocal behavior and on-off switching capabilities across tunable field and frequency ranges. A key innovation of our approach lines in the precise control over the refractive index, enabling adjustments spanning from positive through zero to significantly negative values. This breakthrough has the potential to revolutionize the design and functionality of optical and magnonic devices, offering unprecedented levels of manipulation and performance.

The compatibility of our system with current CMOS circuit technology, along with its simplicity, significantly enhances its practicality for easy incorporation into a wide array of electronic and photonic devices. Additionally, the planar configuration of our PMH system marks a departure from traditional, complex designs, playing a crucial role in miniaturizing quantum hybrid structures. This approach enables the realization of materials with NRI in the beyond-terahertz-range, under effective magnetic field control.

This groundbreaking work opens a new era in photonics and magnonics, setting a foundation for the development of ultra-compact, high-frequency devices. Additionally, it paves the way for further advancements in quantum hybrid systems, promising a future abundant in technological innovations and superior device functionalities.

## Methods

**Sample Fabrication** The hybrid sample used in this study is depicted in Fig. 1b. A YIG film (green color) was placed on top of the 50 Ω characteristic impedance microstrip line on the front side of the sample, while an ISRR was fabricated on the ground plane just below the microstrip line on the backside. The dimensions of the ISRR, microstrip line, and YIG film used in this study are depicted in Fig. S3. The dimensions of the ISRR were determined through simulation using CST-STUDIO. The samples of the microstrip line and the ISRR on the ground plane were fabricated using a conventional photolithography process.

**Scattering parameters Measurement** The measurement setup used in this study, as illustrated in Fig. 1a, involved exciting and detecting the dynamic modes of the YIG film by applying microwave AC currents along the microstrip line and coupling with the electrodynamic photon mode of the ISRR. The input and output of the microstrip feeding line were connected to a two-port vector network analyzer (VNA, Agilent PNA series E8362C) through microwave connectors. The entire ISRR-YIG hybrid sample was subjected to a DC bias magnetic field at room temperature. S-parameters were measured experimentally as a function of the frequency ($f_{AC} = \omega/2\pi$) of AC currents flowing along the microstrip line for different strengths of $\mu_0 H$ applied at an angle $\psi = 33°$ with respect to the x-axis (microstrip line). The angle $\psi$ was chosen because, at this angle, all excited spin-wave modes exhibit a zero-group velocity ($v_g = 0$). This configuration results in a singular FMR line without subsidiary lines from non-zero wave number modes that would occur at other field angles, thereby simplifying the NRI analysis and reducing complexity [51]. To isolate the S-parameters of the ISRR/YIG hybrid in the region of interest, the background signal of the empty microstrip line was subtracted from the entire signal measured from the whole sample [28].

## Data availability

The data presented in this study are available within the article and its Supplementary Information files or from the corresponding author upon request.

## References

[1] Huebl, H., Zollitsch, C. W., Lotze, J., Hocke, F., Greifenstein, M., Marx, A., Gross, R. & Goennenwein, S. T. B. *Phys. Rev. Lett.* **111**, 127003 (2013).

[2] El-Ganainy, R., Makris, K. G., Khajavikhan, M., Musslimani, Z. H., Rotter, S. & Christodoulides, D. N. Non-Hermitian physics and PT symmetry. *Nat. Phys.* **14**, 11–19 (2018).

[3] Miri, M.-A. & Alù, A., Exceptional points in optics and photonics. *Science* **363**, 7709 (2019).

[4] Wang, C., Sweeney, W. R., Stone, A. D. & Yang, L. Coherent perfect absorption at an exceptional point. *Science* **373**, 1261-1265 (2021).

[5] Zhang, D., Luo, X.-Q., Wang, Y.-P., Li, T.-F. & You, J. Q. Observation of the exceptional point in cavity magnon-polaritons. *Nat. Commun*. **8**, 1368 (2017).

[6] Yang, Y., Wang, Y.-P., Rao, J. W., Gui, Y. S., Yao, B. M., Lu, W. & Hu, C.-M. Unconventional singularity in anti-parity-time symmetric cavity Magnonics. *Phys. Rev. Lett.* **125**, 147202 (2020).

[7] Qian, J., Meng, C. H., Rao, J. W., Raw, Z. J., An, Z., Gui, Y. & Hu, C. -M. Non-Hermitian control between absorption and transparency in perfect zero-reflection magnonics. *Nat Commun* **14**, 3437 (2023).

[8] Han, Y., Meng, C., Pan, H., Qian, J., Rao, Z., Gui, Y. & Hu, C. -M. Bound chiral magnonic polariton states for ideal microwave isolation. *Sci. Adv.* **9**, 4730 (2023).

[9] Kimble, H. J. The quantum internet. *Nature* **453**, 1023 (2008).

[10] Tabuchi, Y., Ishino, S., Noguchi, A., Ishikawa, T., Yamazaki, R., Usami, K. & Nakamura, Y. Coherent coupling between a ferromagnetic magnon and a superconducting qubit. *Science* **349**, 405-408 (2015).

[11] Quirion, D. L., Tabuchi, Y., Ishino, S., Noguchi, A., Ishikawa, T., Yamazaki, R. & Nakamura Y. Resolving quanta of collective spin excitations in a millimeter-sized ferromagnet. *Sci. Adv.* **3**, 1603150 (2017).

[12] Zhang, X., Zou, C.-L., Jiang, L. & Tang, H. X. Cavity magnomechanics. *Sci. Adv.* **2**, 1501286 (2016).

[13] Zhang, X., Zou, C.-L., Zhu, N., Marquardt, F., Jiang, L. & Tang, H. X. Magnon dark modes and gradient memory. *Nat. Commun*. **6**, 8914 (2015).

[14] Yuan, H. Y., Yan, P., Zheng, S., He, Q. Y., Xia, K. & Yung, M.-H. Steady Bell State Generation via Magnon-Photon Coupling. *Phys. Rev. Lett*. **124**, 053602 (2020).

[15] Han, J.-X., Wu, J.-L., Wang, Y., Xia, Y., Jiang, Y.-Y. & Song, J. Tripartite high-dimensional magnon-photon entanglement in phases with broken. *Phys. Rev. B* **105**, 064431 (2022).

[16] Lachance-Quirion, D., Wolski, S. P., Tabuchi, Y., Kono, S., Usami, K. & Nakamura, Y. Entanglement-based single-shot detection of a single magnon with a superconducting qubit. *Science* **367**, 425-428 (2020).

[17] Wolski, S. P., Lachance-Quirion, D., Tabuchi, Y., Kono, S., Noguchi, A., Usami, K. & Nakamura, Y., Dissipation-Based Quantum Sensing of Magnons with a Superconducting Qubit. *Phys. Rev. Lett*. **125**, 117701 (2020).

[18] Harder, M., Yang, Y., Yao, B. M., Yu, C. H., Rao, J. W., Gui, Y. S., Stamps, R. L. & Hu, C.-M. Level Attraction Due to Dissipative Magnon-Photon Coupling. *Phys. Rev. Lett*. **121**, 137203 (2018).

[19] Bhoi, B., Kim, B., Jang, S.-H., Kim, J., Yang, J., Cho, Y.-J. & Kim, S.-K. Abnormal anticrossing effect in photon-magnon coupling. *Phys. Rev. B* **99**, 134426 (2019).

[20] Boventer, I., Kläui, M., Macêdo, R. & Weides, M. Steering between level repulsion and attraction: broad tunability, *New J. Phys.* **21,** 125001 (2019).

[21] Wang, Y. P., Rao, J. W., Yang, Y., Xu, P. C., Gui, Y. S., Yao, B. M., You, J. Q. & Hu, C.-M. Nonreciprocity and Unidirectional Invisibility in Cavity Magnonics, *Phys. Rev. Lett*. **123**, 127202 (2019).

[22] Zian, J., Rao, J.W., Gui, Y.S., Wang, Y.P., An, Z.H. & Hu, C.-M. Manipulation of the zero-damping conditions and unidirectional invisibility in cavity magnonics, *Appl. Phys. Lett.* **116**, 192401 (2020).

[23] Vesselago, V. G. The Electrodynamics of substances with simultaneously negative values of ε and μ, *Sov. Phys. Usp.* **10**, 509 (1968).

[24] Smith, D. R., Padilla, W. J., Vier, D. C., Nemat-Nasser, S. C. & Schultz, S. Composite medium with simultaneously negative permeability and permittivity, *Phys. Rev. Lett*. **84**, 84 (2000).

[25] Shelby, R. A., Smith, D. R. & Schultz, S. Experimental verification of a negative index of refraction, *Science* **292**, 77-79 (2001).

[26] Fleury, R., Sounas, D. L. & Alù, A. Negative Refraction and Planar Focusing Based on Parity-Time Symmetric Metasurfaces. *Phys. Rev. Lett*. **113**, 023903 (2014).

[27] Bhoi, B., Kim, B., Kim, J., Cho, Y.-J. & Kim, S.-K. Robust magnon-photon coupling in a planar-geometry hybrid of inverted split-ring resonator and YIG film, *Sci. Rep.* **7**, 11930 (2017).

[28] Bilzer, C., Devolder, T., Crozat, P., Chappert, C., Cardoso, S. & Freitas, P. P. Vector network analyzer ferromagnetic resonance of thin films on coplanar waveguides: Comparison of different evaluation methods, *J. Appl. Phys*. **101**, 074505 (2007).

[29] Metelmann, A. & Clerk, A. A. Nonreciprocal Photon Transmission and Amplification via Reservoir Engineering. *Phys. Rev. X* **5**, 021025 (2015).

[30] Yao, B., Yu, T., Zhang, X., Lu, W., Gui, Y., Hu, C.-M. & Blanter, Y. M. The microscopic origin of magnon-photon level attraction by traveling waves: Theory and experiment. *Phys. Rev. B* **100**, 214426 (2019).

[31] Nicolson, A. M. & Ross, G. F. Measurement of the intrinsic properties of materials by time-domain techniques*, IEEE Trans. Instr. Meas*. **19**, 377-382 (1970).

[32] Weir, W. B. Automatic measurement of complex dielectric constant and permeability at microwave frequencies. *Proceedings of the IEEE* **62**, 33–36 (1974).

[33] Chen, X., Grzegorczyk, T. M., Wu, B.-I., Pacheco, J., Jr. & Kong, J. A. Robust method to retrieve the constitutive effective parameters of metamaterials, *Phys. Rev. B* **70**, 016608 (2004).

[34] McCall, M. W., Lakhtakia, A. & Weiglhofer, W. S. The negative index of refraction demystified, *Eur. J. Phys*. **23**, 353 (2002).

[35] Lakhtakia, A., McCall, M. W. & Weiglhofer, W. S. Brief overview of recent developments on negative phase–velocity mediums (alias Left–handed materials), *AEU - Int. J. Electron.*

*Commun*. **56**, 407-410 (2002).

[36] Depine, R. A. & Lakhtakia, A. A new condition to identify isotropic dielectric-magnetic materials displaying negative phase velocity, *Micro. Opt. Tech. Lett*. **41**, 315 (2004).

[37] Kaina, N., Lemoult, F., Fink, M. & Lerosey, G. Negative refractive index and acoustic superlens from multiple scattering in single negative metamaterials, *Nature* **525**, 77 (2015).

[38] Alù, A., Engheta, N., Erentok, A. & Ziolkovwski, R. W. Single-negative, double-negative, and low-index metamaterials and their electromagnetic applications, *IEEE Ant. Prog. Mag.* **49**, 23-36 (2007).

[39] Tung, N. T., Lam, V. D., Park, J. W., Cho, M. H., Rhee, J. Y., Jang, W. H. & Lee, Y. P. Single- and double-negative refractive indices of combined metamaterial structure, *J. Appl. Phys*. **106**, 053109 (2009).

[40] Hsu, C., Zhen, B., Stone, S. A., Joannopoulos, J. D. & Soljacic, M. Bound states in the continuum. *Nat. Rev. Mater.* **1**, 16048 (2016).

[41] Kang, M., Liu, T., Chan, C. T. & Xiao, M. Applications of bound states in the continuum in photonics. *Nat. Rev. Phys*. **5**, 659–678 (2023).

[42] Pozar, D. M. *Microwave Engineering 4$^{th}$ edition*, (Wiley, New Jersey, 2011).

[43] Caloz, C., Itoh, T. *Electromagnetic Metamaterials: Transmission Line Theory and Microwave Applications*. (Wiley, New Jersey, 2005).

[44] Bonache, J., Gil, M., Gil, I., Garcia-Garcia, J. & Martin, F. On the electrical characteristics of complementary metamaterial resonators. *IEEE Microw. And Wirel. Compon. Lett.*, **16**, 543-545 (2006).

[45] Polder, D. On the Quantum Theory of Ferromagnetic Resonance. *Phys. Rev.* **73**, 1116 (1948).

[46] Rachford, F. J., Armstead, D. N., Harris, V. G. & Vittoria, C. Simulations of ferrite-dielectric-wire composite negative index materials, *Phys. Rev. Lett*. **99**, 057202 (2007).

[47] Zhao, H., Zhou, J., Kang, L. & Zhao, Q. Tunable two-dimensional left-handed material consisting of ferrite rods and metallic wires, *Opt. Exp*. **17**, 13373-13380 (2009).

[48] Mao, S.-G., Wu, M.-S., Chueh, Y.-Z. & Chen, C. Modeling of symmetric composite right/left-handed coplanar waveguides with applications to compact bandpass filters, *IEEE*

*Trans. Microw. Theory Tech*. **53**, 3460-3466 (2005).

[49] Xu, H.-X., Wang, G.-M., Qi, M.-Q. & Xu, Z.-M. Theoretical and experimental study of the backward-wave radiation using resonant-type metamaterial transmission lines. *J. Appl. Phys*. **112** 104513 (2012).

[50] Liu, R., Cui, T. J., Huang, D., Zhao, B. & Smith, D. R., Description and explanation of electromagnetic behaviors in artificial metamaterials based on effective medium theory, *Phys. Rev. E* **76**, 026606 (2007).

[51] Kim, B., Bhoi, B. & Kim, S-K. Spin-wave excitation and critical angles in a hybrid photon-magnon-coupled system. *J. Appl. Phys.* **126**, 163902 (2019).

## Acknowledgments

This research was supported by the Basic Science Research Program through the National Research Foundation of Korea (NRF) funded by the Ministry of Science, ICT & Future Planning (NRF-2021R1A2C2013543). The Institute of Engineering Research at Seoul National University provided additional research facilities for this work.

## Author contributions

S.-K.K. B.K. conceptualized the idea and designed the project. J.K., B.-J.K., and H.J. prepared the samples and conducted the experiments, with B.K. providing essential support. J.K. developed the analytical circuit model under the guidance of B.K. and S.-K. K, in collaboration with B.-J.K. S.-K.K. led and supervised the project.  S.-K.K., J.K., and B.K wrote the manuscript, with all authors participating in providing feedback to refine and edit the manuscript.

**Correspondence and requests for materials** should be addressed to S.-K. K. (sangkoog@snu.ac.kr).

# Figure Legends

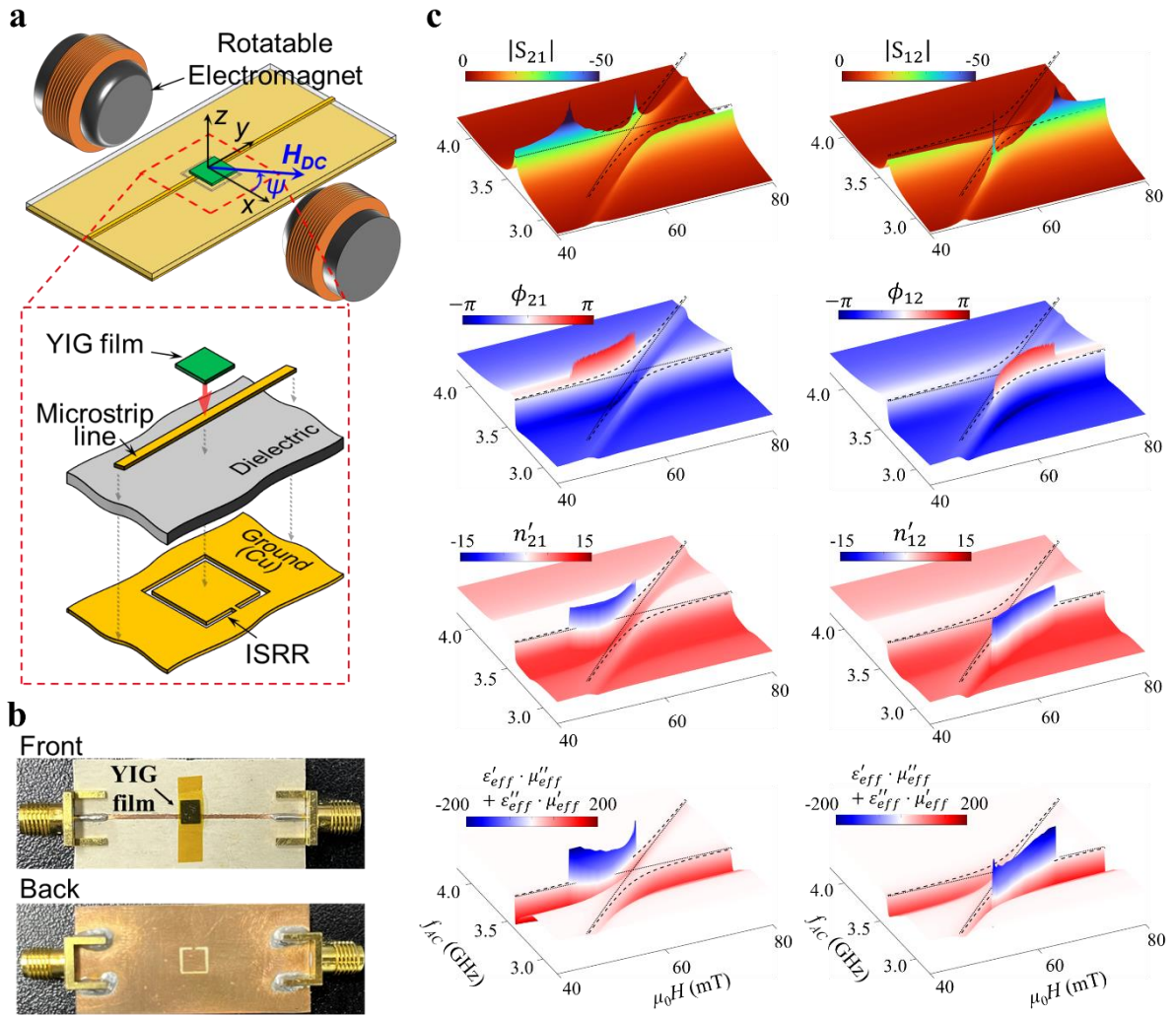

**Fig. 1 | Scattering parameter measurement and NRI estimation in an ISRR/YIG hybrid sample**. **a.** Setup schematic for measuring $S_{ij}$ scattering parameters versus AC current frequency ($f_{AC} = \omega/2\pi$) and varying magnetic field strengths ($\mu_0 H$), applied at a 33° angle to the *x*-axis. **b.** Photo images of the ISRR/YIG sample, showcasing both front and back side views. **c.** Experimental data presentation: Magnitude (dB scale, top row) and phase (radian, second row) for $S_{21}$ (left) and $S_{12}$ (right) transmission as functions of $f_{AC}$ and $\mu_0 H$. Refractive index (third row) estimated from the scattering parameters, is highlighted in blue. The calculation of $\varepsilon'_{eff} \cdot \mu''_{eff} + \varepsilon''_{eff} \cdot \mu'_{eff}$ (last row) should be negative to indicate the occurrence of NRI. Solid

lines trace the uncoupled photon and magnon modes ($\omega_{ISRR}/2\pi$ and $\omega_r/2\pi$), with dashed lines indicating the split hybrid modes from photon-magnon interactions, as described by Eq. 1.

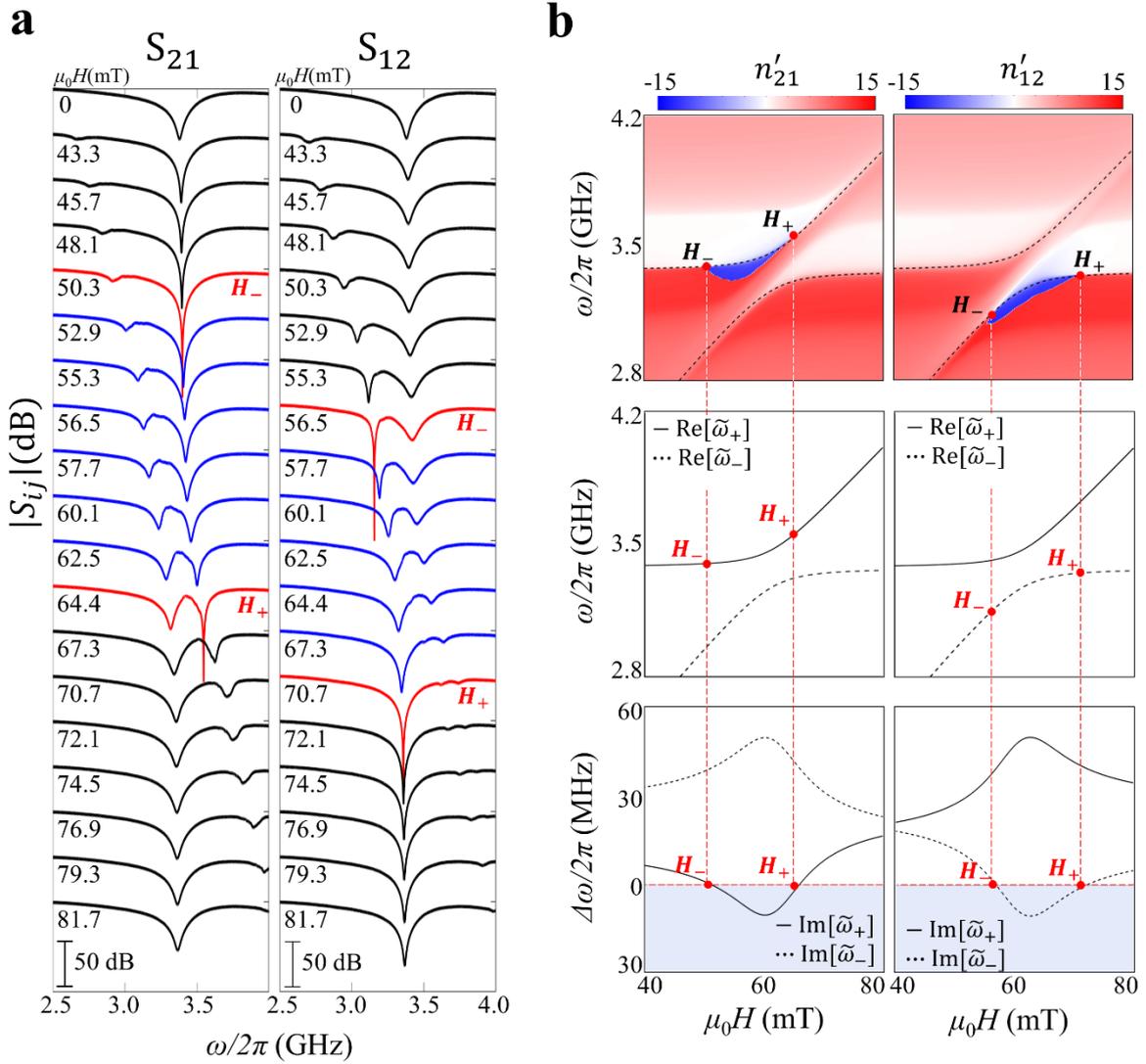

**Fig. 2** | **Experimental confirmation of zero and negative damping in a non-Hermitian sample. a.** Line profiles of |S$_{21}$| and |S$_{12}$| against frequency ($\omega/2\pi$) at specific magnetic fields. Narrow linewidths marked in red indicate zero damping, while the blue-colored intermediate region denotes negative damping (anti-damping). **b.** Analysis of the photon-magnon hybrid modes: The top panel shows the refractive index, the middle panel details real eigenvalues, and the bottom panel focuses on imaginary eigenvalues for $\kappa_c = 0.0296 + 0.0088i$ (left) and $\kappa^*_c = 0.0296 - 0.0088i$ (right), according to Eq. 1. Solid and dashed lines trace the eigenvalues of the higher- and lower-branch modes, respectively. Red dash lines denote zero linewidths, and the blue area represents regions of negative linewidth (indicative of negative

damping). The vertical lines at field strengths, denoted as $H_-$ and $H_+$, highlight the points of zero damping and delineate the range within which NRI is observable.

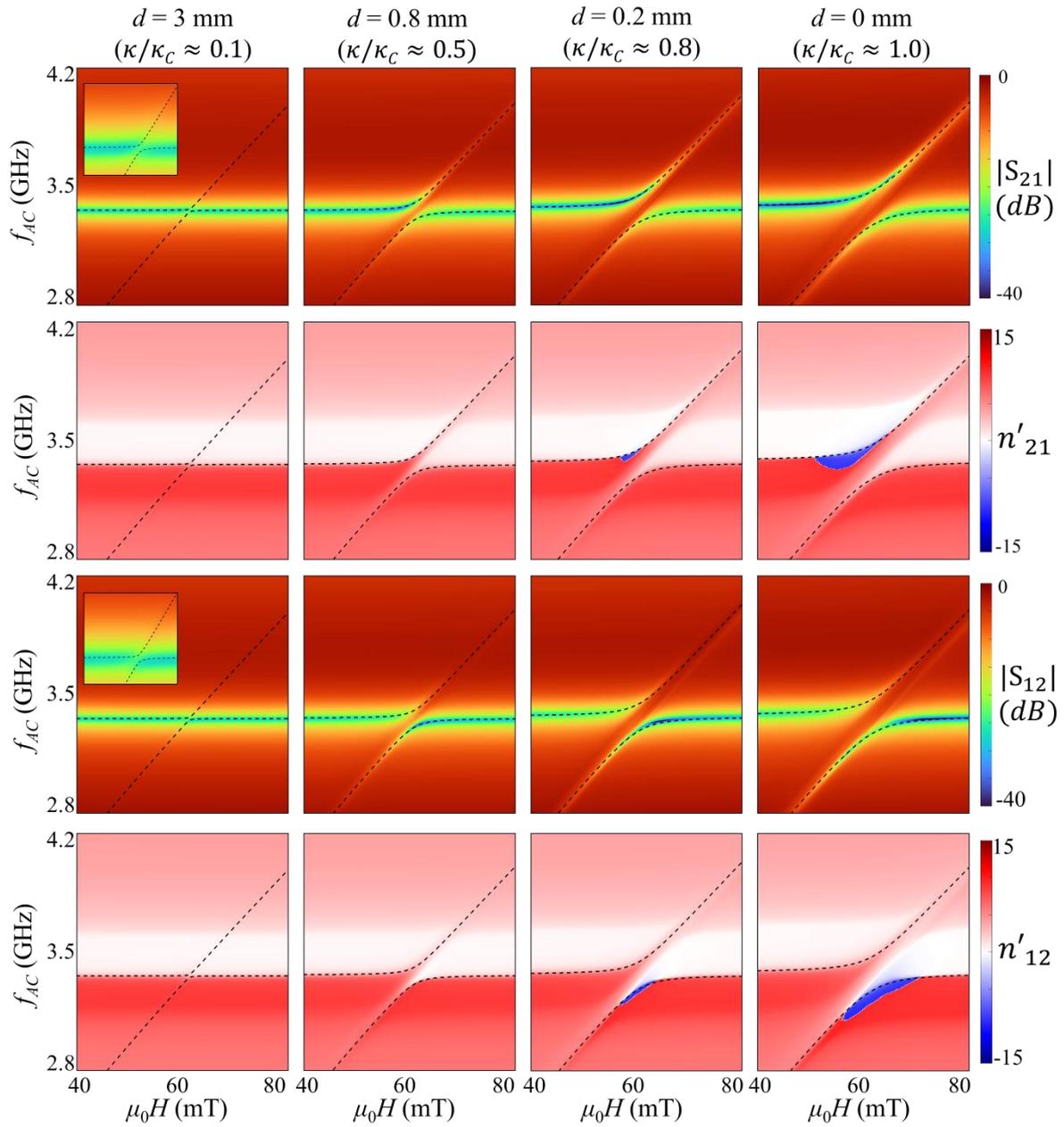

**Fig. 3 | Coupling strength influence on $S_{21(12)}$ and $n'$ measurements.** $|S_{21}|$ spectra (first row) and $|S_{12}|$ spectra (third row), along with the derived refractive indices $n'_{21}$ (second row) and $n'_{12}$ (forth row), plotted in a range of 2.8 GHz $< f_{AC} <$ 4.2 GHz and 40 mT $< \mu_0 H <$ 80 mT. These measurements correspond to varied coupling strengths, $\kappa/\kappa_c = 0.1, 0.5, 0.8$, and 1.0, based on fitting to the higher and lower hybrid modes. The parameter $d$ indicates scotch tape thickness, affecting the coupling constant $\kappa$ with $\kappa_c$ (at $d = 0$). Insets show the coupled mode for $d = 3$ mm in ranges of 3.3 GHz $< f_{AC} <$ 3.5 GHz and 55 mT $< \mu_0 H <$ 65 mT.

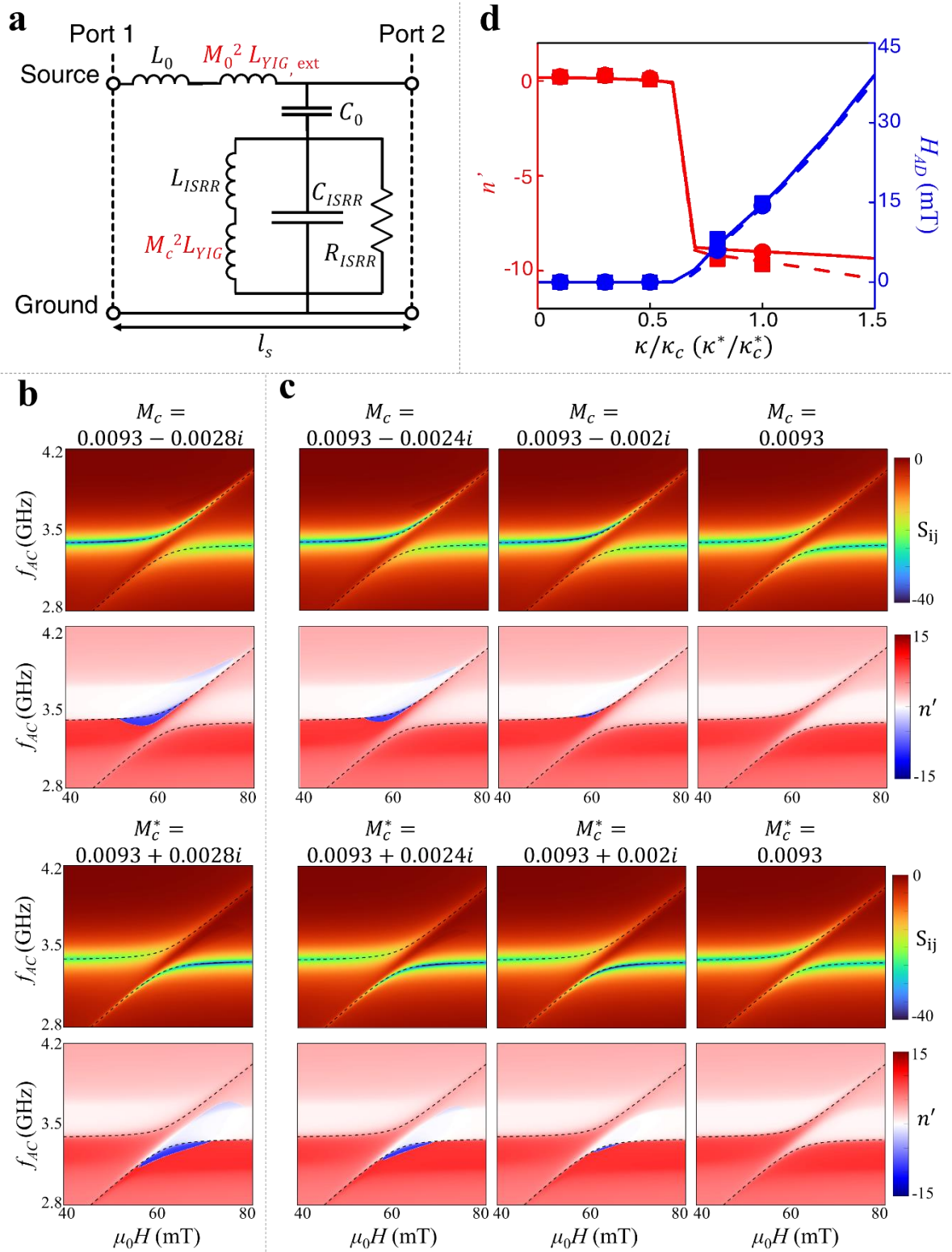

**Fig. 4 | Circuit model representation and numerical analysis of the ISRR/YIG sample. a.** A circuit model based on the transmission line theory for coupled ISRR/YIG hybrid, illustrating the theoretical framework for the subsequent analyses. **b.** Numerical calculations

for $S_{21}$ (top) and $S_{12}$ (bottom) spectra, along with the derived real part of the refractive index $n'$ for coupling constants $M_c = 0.0093 - 0.0028i$ and $M^*_c = 0.0093 + 0.0028i$. Dashed lines indicate the split hybrid modes ($\omega_\pm/2\pi$) derived from Eq. 1. **c.** Circuit model analysis of varying imaginary parts of the coupling constant ($\text{Im}[M_c]$) on $|S_{ij}|$ and $n'$, for specific values, while maintaining the real part constant ($M_c = 0.0093$), displaying the dependence of spectral and refractive properties on the imaginary coupling constant, where (top) $\text{Im}[M_c] = -0.0024, -0.002, 0$ and (bottom) $\text{Im}[M_c] = 0.0024, 0.002, 0$. **d.** Numerical calculation of the peak value of $n'$ (red) and the field region $\Delta H_{ND}$ indicating anti-damping (blue) as a function of $\kappa/\kappa_c$ (solid lines) and $\kappa^*/\kappa^*_c$ (dotted lines), comparing with experimental measurements (solid square and circles, respectively), to validate the model against observed NRI phenomena.